\def \SAIT #1 #2 {{\em Mem.\ Soc.\ Astron.\ It.\/} {\bf #1}, #2}
\def \MESS #1 #2 {{\em The Messenger\/} {\bf #1}, #2}
\def \ASTRNACH #1 #2 {{\em Astron. Nach.\/} {\bf #1}, #2}
\def \AAP #1 #2 {{\em Astron. Astrophys.\/} {\bf #1}, #2}
\def \AAL #1 #2 {{\em Astron. Astrophys. Lett.\/} {\bf #1}, L#2}
\def \AAR #1 #2 {{\em Astron. Astrophys. Rev.\/} {\bf #1}, #2}
\def \AAS #1 #2 {{\em Astron. Astrophys. Suppl. Ser.\/} {\bf #1}, #2}
\def \AJ #1 #2 {{\em Astron. J.\/} {\bf #1}, #2}
\def \ANNREV #1 #2 {{\em Ann. Rev. Astron. Astrophys.\/} {\bf #1}, #2}
\def \APJ #1 #2 {{\em Astrophys. J.\/} {\bf #1}, #2}
\def \APJL #1 #2 {{\em Astrophys. J. Lett.\/} {\bf #1}, L#2}
\def \APJS #1 #2 {{\em Astrophys. J. Suppl.\/} {\bf #1}, #2}
\def \APSS #1 #2 {{\em Astrophys. Space Sci.\/} {\bf #1}, #2}
\def \ASR #1 #2 {{\em Adv. Space Res.\/} {\bf #1}, #2}
\def \BAIC #1 #2 {{\em Bull. Astron. Inst. Czechosl.\/} {\bf #1}, #2}
\def \JSQRT #1 #2 {{\em J. Quant. Spectrosc. Radiat. Transfer\/} {\bf #1}, #2}
\def \MN #1 #2 {{\em Mon. Not. R. Astr. Soc.\/} {\bf #1}, #2}
\def \MEM #1 #2 {{\em Mem. R. Astr. Soc.\/} {\bf #1}, #2}
\def \PLR #1 #2 {{\em Phys. Lett. Rev.\/} {\bf #1}, #2}
\def \PASJ #1 #2 {{\em Publ. Astron. Soc. Japan\/} {\bf #1}, #2}
\def \PASP #1 #2 {{\em Publ. Astr. Soc. Pacific\/} {\bf #1}, #2}
\def \NAT #1 #2 {{\em Nature\/} {\bf #1}, #2}
\def \gh {\widehat{g} (r)}
\def \xih {\widehat{\xi} (r)}
\newcommand{\mincir}{\raise -2.truept\hbox{\rlap{\hbox{$\sim$}}\raise5.truept
\hbox{$<$}\ }}

\documentstyle{memsait}
\input epsf.sty
\begin{opening}
\title{LIKELIHOOD ANALYSIS OF THE CfA2+SSRS2 AND OF 
THE LAS CAMPANAS REDSHIFT
SURVEYS}
\author{EMILIA PALLADINO$^1$, LUCA AMENDOLA$^1$}
\institute{$^1$Osservatorio Astronomico di Roma - Viale del Parco Mellini,84, 00136, Roma, Italia}
\date{} 
\end{opening}

\begin{document}

\oddpagefooter{}{}{} 
\evenpagefooter{}{}{} 
\ 
\bigskip

\begin{abstract}
We present the results from the correlation analysis of
two galactic redshift surveys, the extended CfA2+SSRS2 and the slice
centered on $\delta=-12^{\circ}$ of Las Campanas. Furthermore,
we evaluate the likelihood
confidence regions for the CDM model and for the fractal model
parameters. Our results indicate that,
although the CDM model is in good agreement with the data, the fractal
description cannot be ruled out, because of its intrinsic high variance.
\end{abstract}

\vspace{0.5cm}

The availability in the late years of deeper and more accurate redshift
surveys allows us 
not only to better estimate the statistical properties
of the galaxy distribution,
but also the parameters of the theoretical models. Here
we compare the real data with two models that represent
two opposite visions of the luminous matter 
distribution.
On one hand the CDM model and its variants state the
homogeneity of the galaxy distribution at a scale of order of
tens of Mpc (Davis 1997; Cappi et al. 1998); on the other hand
the fractal geometrical description of the Universe (Pietronero et al. 1997),
assumes that it is
completely inhomogeneus at all scales.

To test if a particle distribution has fractal properties, we define the
statistical estimator of the correlation function
$g(r)=1+ \xi (r)$,
where $\xi(r)$ is the standard correlation function:
if the distribution is fractal with dimension $D$,
$g(r)$ decreases as $r^{3-D}$, otherwise it flattens.
The correlation $g (r)$ has the further advantage that 
its estimation in finite volume is simply proportional 
to its universal value (Amendola 1998).

The applied statistical method is based on the use of the integrated
correlation function
$\gh=\int g(r) d^3 r$
because the differential quantity
$g$ is very noisy if it is calculated in low density samples
as in Las Campanas (LCRS - Schectman et al. 1996).
As usual, we use volume limited (VL) samples to avoid the
corrections due to the selection function and the possible dependence
on luminosity of the clustering amplitude;
we estimate the count cell volumes using Monte Carlo
and we ensure that the cell boundaries are
{\em completely internal} to the survey geometry
(Pietronero et al. 1997).
The last condition puts a constraint
on the value of $R_M$, the maximum scale reached via this method,
for it
depends on the shape of the survey slice,
{\it i.e.} on its depth $d$ and its minimum angular opening $\theta$.
In the case of very narrow
slice, as the one of LCRS
($\Delta \delta = 1.5^{\circ }$), and if we use spherical cells,
$R_M$ reaches a few
Mpc, while it is crucial to extend the 
analysis up to scales greater than 50 Mpc/$h$
in order to distinguish between the two
competing models.
To increase $R_M$ 
we use count cells with the same shape as the survey slice,
{\it i.e.} same $\theta$ and variable $d$ ({\em
radial} cells - Amendola \& Palladino 1999).
In this way we
reach deeper scales than the spherical window; for example, in the case of
LCRS, we obtain $R_M \approx 200$ Mpc/$h$.

We apply the method to seven VL samples,
four extracted from the slice centered on $\delta=-12^{\circ}$ of LCRS
and three from the extended CfA2+SSRS2 (da Costa et al. 1995).
In Tab. I we summarize the characteristic numbers of the samples. 

To estimate the variance of $\gh$ let us make two assumption: 
the 3-point correlation function is given by the scaling
relation 
$\varsigma_{ijk}=Q(\xi_{ij} \xi_{jk}+\xi_{ij} \xi_{ik}+\xi_{ik} \xi_{jk})$, where 
we put $Q=1$; $\xi_{ij}$ can be approximated by a power law.
So it is possible to demonstrate that the following relation holds (Amendola 1998):
\begin{equation}
\sigma _{\widehat{g}}^{2}=N_{c}^{-1}\left[ N_{0}^{-1}\left( 1+\widehat{\xi }%
\right) +\sigma ^{2}\left( 1+2\widehat{\xi }\right) \right],
\label{vargh}
\end{equation}
where $N_c$ is the number of independent cells 
and $N_0$ is the number of galaxies contained on average in the cells.

The expression
$P(k)=A\,k\,T^{2}\left( \Gamma ,k\right) \,G\left( \Omega _{m},\Omega
_{\Lambda },\sigma _8, \sigma _{v}\right)$
is the power spectrum in a CDM flat universe,
where $T \left( \Gamma ,k\right)$ is the transfer function
of Bardeen et al. (1986),
$G$ includes the redshift and nonlinear corrections
(Kaiser 1987; Peacock \& Dodds,
1996), $\sigma _{v}$ is the line-of-sight velocity dispersion and the
subscripts $\Lambda$ and $m$ refer to cosmological constant and total matter, respectively.
For the fractal model, the usual integrated correlation function with
dimension $D$ and normalization $r_0$ is given by 
$\xih = \left( r / r_0 \right) ^{D-3} -1$.
Using these quantities it is possible to evaluate Eq. \ref{vargh},
remembering that $\xi (r)$ is the Fourier transform of $P (k)$ and
$\xih$ its volume integral.
Notice that in the relation
$\sigma ^{2}=\left( 2\pi ^{2}\right) ^{-1}\int P\left( k\right) W_{c}\left(
k\right) k^{2}dk$
the window $W_{c} \equiv W^2$ only if it is spherical, otherwise it has
to be calculated numerically. 

Finally we perform the likelihood analysis including the free parameters
of the model both in the mean {\em and} in the variance.
For CDM these are the shape factor $\Gamma$ and the galaxy normalization $\sigma_{8}$, 
where we have fixed
$\sigma_{v}=300$ km sec$^{-1}$,
$\Omega_m =0.4$ and $\Omega_\Lambda =0.6$; for the fractal model the parameter
is $D$ while $r_0$
is fixed to be equal to the maximum depth
with respect all the samples, {\it i.e.} 437 Mpc/$h$.
The results of the parameters estimation are given in Tab. I.

\begin{table}[h]
\begin{tabular}{|l|c|c|c|c|c|c|l|c|c|c|c|c|c|}
\hline
\multicolumn{7}{|c|}{Las Campanas}      &\multicolumn{7}{|c|}{CfA2+SSRS2}  \\
\hline
VL    &$\Delta R$  &$N_{gal}$ &$\langle M_S \rangle$ & $\Gamma$  &$\sigma_{8}$     & $D$ &VL      &$R$  &$N_{gal}$ &$M_{lim}$  & $\Gamma$  &$\sigma_{8}$ & $D$       \\
\hline 
lc410   & 130        & 510      & $-21.3$              &0.2       &0.9               &2.3  &cs19       & 80  & 840      & $-19.0$     & 1.0       & 1.7          & 2.2      \\ 
lc330   & 140        & 840      & $-20.8$              &0.1       &0.9               &2.4  &cs20       & 125 & 492      & $-20.0$     & 0.2       & 0.7          & 2.8      \\
lc297   & 150        & 818      & $-20.4$              &0.2       &0.9               &2.4  &cs205      & 160 & 212      & $-20.5$   & 0.1       & 1.3          & 2.6      \\
lc437   & 213        & 492      & $-21.3$              &0.2       &0.8               &2.5  &           &     &          &           &           &              &          \\
\hline
\end{tabular}
\caption{{\bf VL samples and Likelihood results} - LCRS has $\Delta \alpha \approx 80^{\circ}$ 
and $\Delta \delta = 1.5^{\circ}$ and because it has two limiting magnitudes, 
each VL has two cuts in distance, so $\Delta R = R_2 - R_1$.
We cut CfA2+SSRS2 to have a regular slice with $\Delta \alpha \approx 80^{\circ}$ 
and $\Delta \delta \approx 45^{\circ}$.}
\end{table}

\begin{figure}
\epsfxsize=10 cm 
\hspace{3.5cm}
\epsfbox{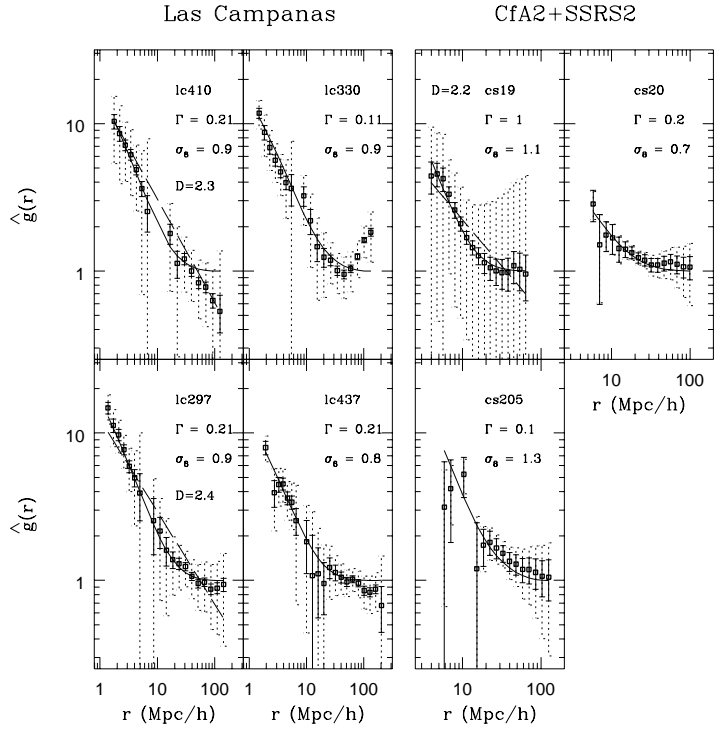}
\caption[h]{ - $\gh$ in all the seven VL analyzed
with the errors expected in a CDM and in a fractal model ({\em dotted lines}).
The parameters used to estimate the error bars are given in Tab. II}
\label{gtrend}
\end{figure}

\begin{figure}
\epsfysize=5cm 
\hspace{2.5cm}
\epsfbox {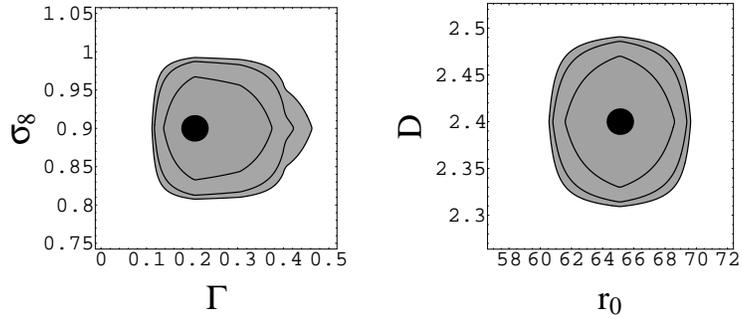}
\caption[h]{ - Here are the contour plots (68, 90, 95\%) of the product of
the likelihood functions for each VL of LCRS; on the left
there is the CDM model, on the right the fractal one.}
\label{contplotlc}
\end{figure}

\begin{figure}
\epsfysize=5cm 
\hspace{2.5cm}
\epsfbox {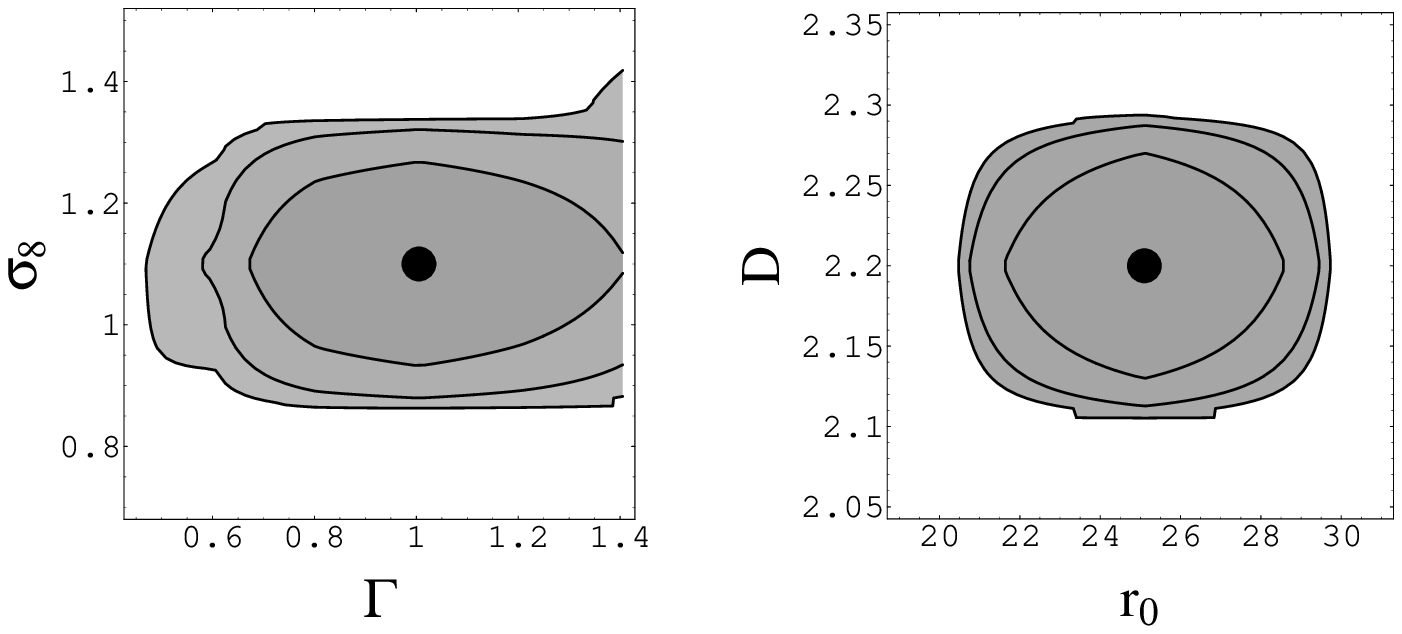}
\caption[h]{ - The contour plots (68, 90, 95\%) of the VL cs19 of
CFA2+SSRS2; on the left
there is the CDM model, on the right the fractal one.}
\label{contplotcs19}
\end{figure}

As we can see in Fig. \ref{gtrend} the scales we
reach are the largest ever reached in $\gh$ statistics.
In Fig. \ref{contplotlc}, we have shown the contour plots of 
the product of the likelihood functions of each VL of
LCRS, because the estimated parameters are similar in all of them.
Our results for LCRS are: $0.12 \mincir \Gamma \mincir 0.46$,
$0.81 \mincir \sigma_8 \mincir 0.99$
for CDM; $2.31 \mincir D \mincir 2.49$ for fractal. 
The estimated parameters of CfA2+SSSRS2 are different from one VL to 
the other (see Tab. II for details) so we do not evaluate
the likelihood function product; we show as an example the contour plot
of cs19 (Fig. \ref{contplotcs19}).
Our conclusions are that $\Gamma \simeq 0.2$ in samples that have
around the same absolute magnitude; remembering that
in the CDM model $\Gamma = \Omega_m h$, this agrees with $\Omega_m = 0.4$
and $h = 0.6 \div 0.7$.
Furthermore, we cannot yet reject the fractal model due to its
intrinsic high variance
(Fig. \ref{gtrend});
this shows that it is necessary
to check the validity of a model including its
full variance in the likelihood.

\vspace{0.5cm}
We acknowledge financial support by the Ministry of University and
of Scientific and Technical Research.

\end{document}